%% file: main-hairpin.tex
\documentclass{jfm}
\usepackage{graphicx}
\usepackage{subfigure}
\usepackage{amssymb}
\usepackage{amsmath}
\usepackage{mathrsfs}
\usepackage{placeins}
\usepackage{natbib}
\usepackage{caption3}
\bibpunct{(}{)}{,}{s}{,}{,}

\usepackage[bookmarksnumbered=true,breaklinks=true,colorlinks,citecolor=blue,linkcolor=blue]{hyperref}

\makeatletter

\newcommand{\numtoRoman}[1]{\expandafter\@slowromancap\romannumeral #1@}

\renewcommand{\figurename}{Figure}
\usepackage[usenames]{color}

\newcommand{\revision}[1]{\textcolor{black}{#1}}
\newcommand{\revisionb}[1]{\textcolor{black}{#1}}

\title[Coherent states with hairpin-like structure]{Exact coherent states with hairpin-like vortex structure in channel flow}
\shortauthor{A.~Shekar and M.~D.~Graham}

\author{Ashwin Shekar\aff{1} \and  Michael D.~Graham\aff{1}\aff{2}\corresp{\email{mdgraham@wisc.edu}}}
\affiliation{\aff{1}Department of Chemical and Biological Engineering, University of Wisconsin-Madison, Madison, WI 53706 USA
\aff{2}Kavli Institute for Theoretical Physics, University of California, Santa Barbara, CA 93106 USA}

\date{\today}

\begin{document}

\maketitle

\begin{abstract}
\input{abstract}
\end{abstract}
\section{Introduction}

\input{intro-hairpins}
\input{intro-ECS}
\section{Formulation}
\input{formulation}

\section{Results and Discussion}
\input{narrative-new}

\section{Conclusions}
\input{Conclusions}
\section*{Acknowledgments}
This work was supported by the National Science Foundation through grant CBET-1510291 (Fluid Dynamics Program) and PHY-1125915 (Kavli Institute for Theoretical Physics), and the Air Force Office of Scientific Research through grant FA9550-15-1-0062 (Flow Interactions and Control Program). The authors are grateful to John Gibson for development and distribution of \texttt{ChannelFlow} and to Bruno Eckhardt for helpful discussions. 
\bibliographystyle{jfm}
\bibliography{turbulence-paperslib}
\end{document}

%% file: abstract.tex
Hairpin vortices are widely studied as an important structural aspect of wall turbulence. 
The present work describes, for the first time, nonlinear traveling wave solutions to the Navier--Stokes equations in the channel flow geometry -- exact coherent states (ECS) -- that display \revision{hairpin-like} vortex structure. This solution family comes into existence at a saddle-node bifurcation at Reynolds number $\Rey=666$. At the bifurcation, the solution has a highly symmetric quasistreamwise vortex structure similar to that reported for previously studied ECS.  With increasing distance from the bifurcation, however, both the upper and lower branch solutions develop a vortical structure characteristic of hairpins: a spanwise-oriented ``head'' near the channel centerplane where the mean shear vanishes connected to counter-rotating quasistreamwise ``legs'' that extend toward the channel wall. At $\Rey=1800$, the upper branch solution has mean and Reynolds shear-stress profiles that closely resemble those of turbulent mean profiles in the same domain. 

%% file: intro-hairpins.tex

\begin{figure}
 		\begin{center}
			\subfigure[]
  			{
 				 \includegraphics[width=0.4\textwidth]{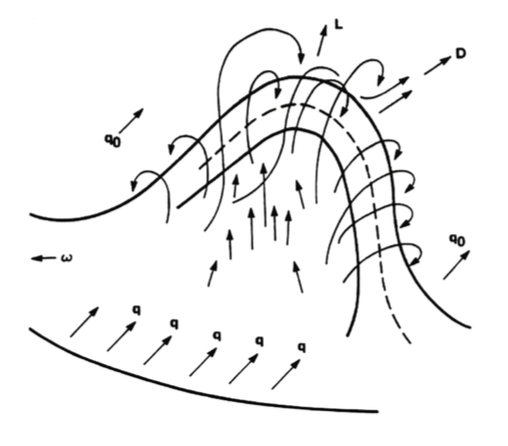} 
  				 \label{fig:Theo_2}
 		 	}
 		 	 \hskip 4ex
 			 \subfigure[]
  			{
 			 \includegraphics[width=0.4\textwidth]{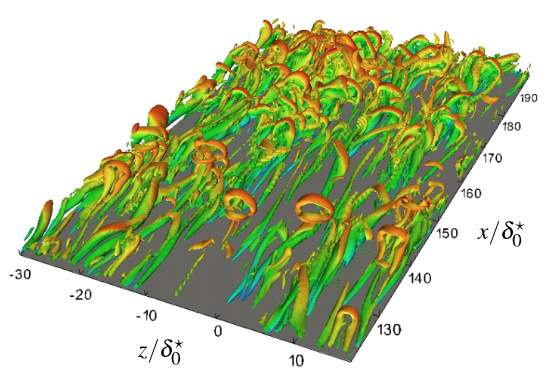} 
  			 \label{fig:Hairpin_Schlatter}
  			}
  			\subfigure[]
  			{
 			 \includegraphics[width=0.4\textwidth]{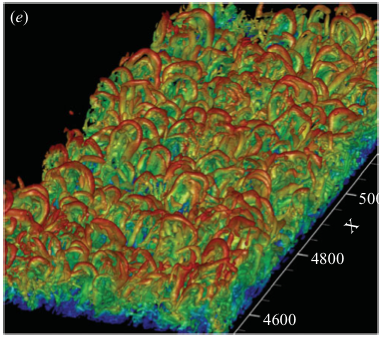}
  			 \label{fig:WuMoinHairpins}
  			}
  			
  			\caption[Hairpins in different contexts]{(a) Theodorsen's conceptualized picture of a hairpin \citep{Theodorsen:1952tn}. \revision{(b) Isosurfaces of $\lambda_{2}$-criterion showing hairpin-like structures during transition in boundary layers\citep{EitelAmor:2015gn}.} (c) Isosurfaces of vortex strength $Q$ illustrating hairpin vortices in direct simulation results for a turbulent boundary layer\citep{WU:2009fn}. 
  			}
  			\label{fig:S1SNB}
		\end{center}
\end{figure}

Ever since the observations of \cite{Theodorsen:1952tn} \revision{(Figure \ref{fig:Theo_2})}, the role of \emph{hairpin} or \emph{horseshoe} vortices in turbulent shear flows has been an important topic in the study of turbulent structure \citep{Robinson:1991ga,Adrian:2007hy,Smits:2011fn}. In a wall-bounded flow,  hairpin vortices are concentrated regions of vorticity whose shape is described by their name, with the ``legs" of the hairpin comprising counter-rotating quasistreamwise vortices tilted upward away from the wall and joined by a ``head'' of primarily spanwise vorticity far from the wall where the mean velocity gradient (and thus mean spanwise vorticity) are small. 
\revision{Hairpin vortices have been observed in many contexts; for example, Figure \ref{fig:Hairpin_Schlatter} shows results from a simulation of transition, illustrating hairpin-like vortices with heads near the edge of the boundary layer.  \cite{Duguet:2012tw}, in a study of transitional boundary layer flow, observe formation of hairpins just above the region where adjacent streamwise streaks ``pinch'' together, and the robustness of this process led the authors to suggest that it is a quasi-equilibrium feature of the flow. Figure \ref{fig:WuMoinHairpins}, from the boundary layer direct numerical simulation of \cite{WU:2009fn} \revision{at transitional Reynolds numbers}, shows isosurfaces of vortex strength that exhibit a ``forest'' of vortices with clear hairpin structure. \cite{schlatter2014near} also studied the presence of hairpins in boundary layers at transitional Reynolds numbers using interesting vortex eduction techniques. In fully developed turbulence, hairpin vortex structures and packets thereof are closely related to the formation of very large-scale uniform momentum zones \cite{Adrian:2007hy}.} Additionally, vortices of hairpin or related shape play a central role in the attached eddy hypothesis \citep{Townsend:1961vi,Perry:1982ue,Woodcock:2015ji}, an important model for the statistical structure of wall turbulence.  The present work describes a traveling wave solution to the Navier--Stokes equations in minimal channel flow that displays hairpin\revision{-like} vortices, demonstrating that such structures exist as autonomous, self-sustaining patterns.

%% file: intro-ECS.tex
	Important insights into turbulent structure, especially near solid surfaces or at low Reynolds number, have been gained in recent years by looking at turbulence through the lens of nonlinear dynamical systems theory, which relates observed dynamics to underlying state space structure. 
	In all of the canonical wall-bounded shear flow geometries (pipe flow, channel flow, plane shear (Couette) flow), families of nonlinear traveling wave solutions to the Navier-Stokes equations (NSE) that govern fluid motion have been discovered \citep{Waleffe:1998wk,Waleffe:2001wu,Waleffe:2003hh,Wang:2007ii,Hof:2004ug,Eckhardt:2007ka,Eckhardt:2008gd,Duguet:2008bs,Wedin:2004ey}.  They have also been found in the asymptotic suction boundary layer flow \citep{Khapko:2013hs}. These solutions are often denoted ``exact coherent states'' (ECS). Related, but more complex states have been found as well, that are not pure traveling waves but rather ``relative periodic orbits'' that are time-periodic modulo a phase shift in one of the translation-invariant spatial directions \citep{Duguet:2008ev}. (A traveling wave is a relative equilibrium -- a state that is time-invariant modulo phase shift in the downstream direction.) In minimal domains at Reynolds numbers near transition the turbulent dynamics have been found to be organized, at least in part, around these traveling wave and relative periodic orbit solutions (see, e.g.~\cite{Gibson:2008eca,Kawahara:2012iu,Park:2014wt}).  The present work exclusively considers traveling wave exact coherent states.

	The basic scenario for appearance of these states is that as $\Rey$ increases, they   emerge in pairs with finite amplitude at  so-called saddle-node bifurcations. The solution with the higher drag is called an ``upper branch" (UB) solution while the other is a ``lower branch" (LB).  
	The lowest Reynolds number at which these solutions appear is slightly below the value where transition is observed in experiments. For channel flow, our focus here, transition occurs at a Reynolds number $\Rey$ based on the laminar centerline velocity of about $1000$, while various families of ECS come into existence once $\Rey\gtrsim 660$ \citep{Park:2014wt,Wall:2016kw}.  Other coherent states can arise through secondary bifurcations off of the states that appear in this way. 

	The spatial structure of the heretofore-discovered ECS is pairs of counter-rotating streamwise vortices, with the corresponding low-and high-speed streaks associated with the advection of the mean shear by the vortices. Letting $x$, $y$ and $z$ be the mean flow, wall-normal and spanwise directions, respectively, and $L_x$ and $L_z$ the spatial periods of the flow in $x$ and $z$, respectively, traveling wave velocity fields \revisionb{$\boldsymbol{u}(x,y,z,t)=\boldsymbol{u}(x-ct,y,z)$ with wavespeed $c$} have generally been sought in one of two invariant subspaces with respect to symmetry in the $x-z$ plane. If $u, v$ and $w$ are the velocity components in $x,y$, and $z$, solutions have been sought that satisfy either  
a simple reflection symmetry across $z=L_z/2$, 
\begin{equation}
	\begin{bmatrix}u  & v & w\end{bmatrix}(x,y,z-L_z/2,t)=\begin{bmatrix}u & v& -w\end{bmatrix}(x,y,-(z-L_z/2),t) \label{eq:reflect}
 \end{equation}	
or a ``shift and reflect'' symmetry, where 
	\begin{equation}
	\begin{bmatrix}u & v& w\end{bmatrix}(x,y,z-L_z/2,t)=\begin{bmatrix}u & v& -w\end{bmatrix}(x+L_x/2,y,-(z-L_z/2),t).\label{eq:shiftreflect}
	\end{equation}
The latter case leads to sinuous vortex structures, while the former leads to varicose or potentially sinucose ones \citep{Waleffe:1997va}. (We address this distinction below.) The former symmetry is also that of an ideal hairpin vortex.  
In the case of channel flow, reflection symmetry across the centerplane of the channel $y=0$ is also often imposed, in which case
\begin{equation}
	\begin{bmatrix}u & v& w\end{bmatrix}(x,y,z)=\begin{bmatrix}u & -v& w\end{bmatrix}(x,-y,z).\label{eq:centerplanesym}
\end{equation}
Finally, a symmetry that has not received much prior attention, but which turns out to be central to the present work, is the ``$xz$-shift'' symmetry condition
\begin{equation}
	\begin{bmatrix}u & v& w\end{bmatrix}(x,y,z)=\begin{bmatrix}u & v& w\end{bmatrix}(x+L_x/2,y,z+L_z/2).\label{eq:xzshift}
\end{equation}
 
	Traveling wave exact coherent states have been computed in the channel flow geometry by a number of authors \citep{Waleffe:2001wu,Waleffe:2003hh,Nagata:2013ha,Park:2014wt,Zammert:2016fk,Rawat:2016ga,Wall:2016kw,Neelavara:2017ed}. None of these have reported a hairpin vortex structure. In the plane Couette flow case, Itano and Generalis \citep{Itano:2009ho} reported an ECS that they described as a hairpin vortex state. (\cite{Gibson:2009kp} also found the same solution family.) Indeed, vortex lines for this state do display a hairpin shape, but in their case the loop or ``head'' of the hairpin is near the wall, where the shear rate is highest. Thus the local vorticity is dominated by the mean shear, which is of course locally oriented in the $z$-direction.  
\revision{In contrast, hairpin vortices in turbulent flows (e.g. as in Figure \ref{fig:WuMoinHairpins}) have a  head that is a localized region of $z$-oriented vorticity in a background of zero or very weak shear as arises near the centerline in channel flow or the outer edge of a boundary layer.}
%
%
Furthermore, inferring existence of hairpin vortices from vortex lines can be problematic: \emph{any} vortex line that crosses a plane of reflection symmetry must do so normally: for example, even a pair of quasistreamwise vortices that satisfies \eqref{eq:reflect} will display hairpin-shaped vortex lines. More generally, \cite{Robinson:1991ga} points out that ``hairpin-shaped vorticity lines are common in any turbulent shear flow, whether or not hairpin-shaped vortices are present''.  Figure 4 of \cite{Robinson:1991ga} illustrates this point. \revision{In our analysis below, we will infer hairpin-like vortex structure from vortex strength $Q$ and direct examination of velocity and vorticity fields.}

	As noted above, ECS play a role in organizing the state space dynamics of turbulent flows. In shear flows whose laminar state is linearly stable (pipe flow, plane Couette flow, channel flow at low Reynolds number), an important role in organizing state space is also played by the boundary between initial conditions that laminarize and those that become turbulent. At long times, trajectories on this boundary approach \emph{edge states} \cite{Skufca:2006iu}. In boundary layer flow, \cite{Cherubini:2011dk} showed the presence of two different edge states, one dominated by streamwise vortex structures and one by transient localized hairpins. 
The study of \cite{Duguet:2012tw} described above also focuses on edge states and indicates that hairpin-vortex-like structures can be found on the laminar-turbulent boundary.


The present work, which focuses on the channel flow geometry, reports for the first time a family of traveling wave exact coherent states that displays a hairpin\revision{-like} vortex structure. Below we describe the bifurcation diagram for this family, its connection to some previously known states, its structure, and initial evidence indicating its relationship to fully turbulent mean profiles. 

%% file: formulation.tex

We consider incompressible Newtonian flow in the plane Poiseuille (channel) geometry, driven with a constant volumetric flux. The $x, y$ and $z$ coordinates are aligned with the streamwise, wall-normal, and spanwise directions, respectively. Periodic boundary conditions are imposed in the $x$ and $z$ directions with fundamental periods $L_x$ and $L_z$, and no-slip conditions are imposed at the walls $y=\pm h$, where $h=L_y/2$ is the half-channel height. Using the half-height $h$ of the channel and the laminar centerline velocity $U_c$ as the characteristic length and velocity scales, respectively, the nondimensionalized Navier-Stokes equations are then given as
\begin{equation}
\frac{\partial \boldsymbol{u}}{\partial t}+\boldsymbol{u}\cdot\nabla \boldsymbol{u}=-\nabla p+\frac{1}{\Rey}\nabla^2\boldsymbol{u}, \quad \nabla \cdot \boldsymbol{u}=0.
\end{equation}
Here, we define the laminar equivalent Reynolds number for the given volumetric flux as $Re_c = U_c h/\nu$, where $\nu$ is the kinematic viscosity of the fluid. We fix the bulk velocity (volumetric flux) $U_b$ at the laminar value $2U_c/3$, so the Reynolds number $\Rey_b$ based on bulk velocity is given by $\Rey_b=2\Rey/3$.  Characteristic inner scales are the friction velocity $u_{\tau}=(\bar{\tau}_w/\rho)^{1/2}$ and the near-wall length scale or wall unit $\delta_{\nu} = \nu/u_{\tau}$, where $\rho$ is the fluid density and $\bar{\tau}_w$ is the time- and area-averaged wall shear stress. As usual, quantities nondimensionalized by these inner scales are denoted with a superscript ``+". The friction Reynolds number is then defined as $Re_{\tau}=u_{\tau}h/\nu=h/\delta_{\nu}$.

Computations are performed using the open source code \texttt{ChannelFlow}, which performs direct simulations as well as computing exact coherent states by Newton-Raphson iteration \citep{Gibson:2012wh}. In this study, we focus on the domain $L_x \times L_y \times L_z = \pi \times 2 \times \pi/2$, the same box size as used in our prior ECS study \citep{Park:2014wt}. Fourier-Chebyshev-Fourier spectral spatial discretization is applied to all variables, using meshes with $N_x \times N_y \times N_z$ collocation points (in $x$, $y$, and $z$).  A typical resolution used is $(N_x, N_y, N_z) = (48, 81, 48)$, which we have verified is sufficient for converged solutions in the Reynolds number range examined here. 

The exact coherent states computed and studied here all satisfy reflection symmetry across $z=L_z/2$, as given by \eqref{eq:reflect}, as well as  across the centerplane of the channel $y=0$, as given by \eqref{eq:centerplanesym}.  
The solutions of central interest here additionally satisfy the $xz$-shift symmetry given by \eqref{eq:xzshift}.
There are of course other channel flow ECS that do not obey these symmetries (e.g.~\cite{Waleffe:2001wu,Waleffe:2003hh,Nagata:2013ha,Gibson:2014jr,Neelavara:2017ed}). All turbulent trajectories reported here are computed without imposing any symmetries on the flow field.

%% file: narrative-new.tex
\begin{figure} 
 		\begin{center}
			\subfigure[]
  			{
 				 \includegraphics[width=0.45\textwidth]{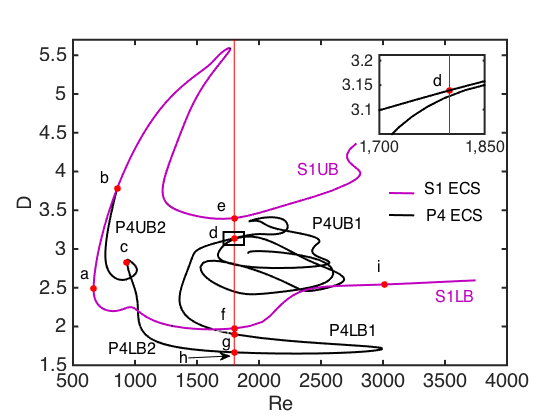}
  				 \label{fig:DRE}
 		 	}
 		 	 \hskip 0.5ex
 			 \subfigure[]
  			{
 			 \includegraphics[width=0.45\textwidth]{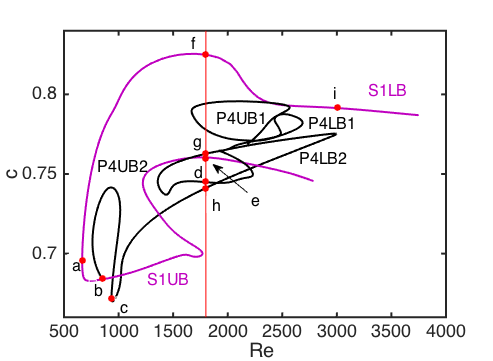}
  			 \label{fig:WaveSpeed}
  			}
  			\caption[Bifurcation diagram]{Bifurcation diagram in terms of (a) dissipation rate $D$ \emph{vs.}~$\Rey$ and (b) wave speed $c$ \emph{vs.}~$\Rey$. Inset in (a) is a zoom-in of the box close to point \textbf{d}.}
  			\label{fig:P4bifdiagram}
		\end{center}
\end{figure}
\begin{figure}
 		\begin{center}
			\subfigure[]
  			{
 				 \includegraphics[width=0.45\textwidth]{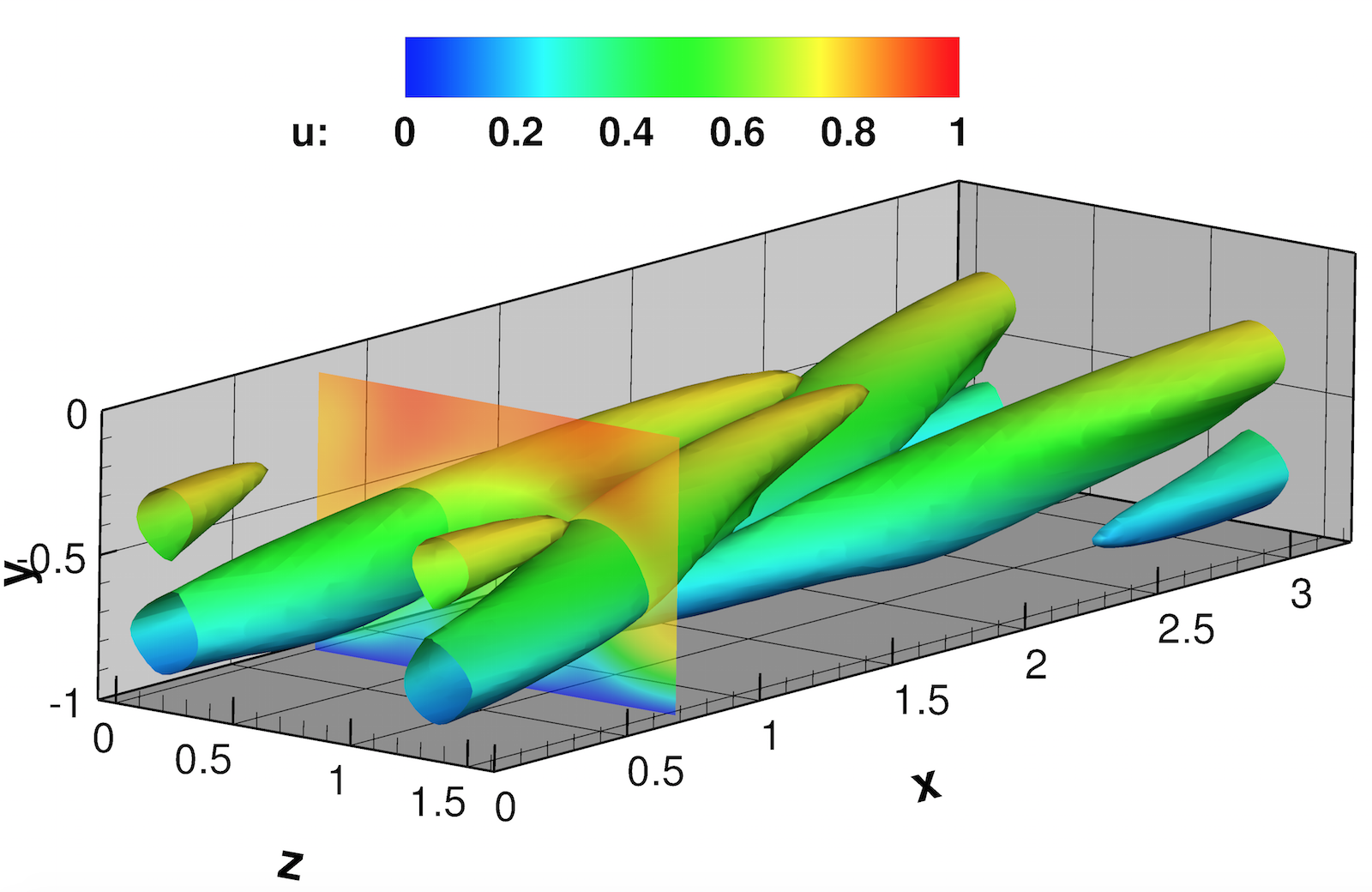} 
  				 \label{fig:SNB}
 		 	}
 		 	 \hskip 4ex
 			 \subfigure[]
  			{
 			 \includegraphics[width=0.45\textwidth]{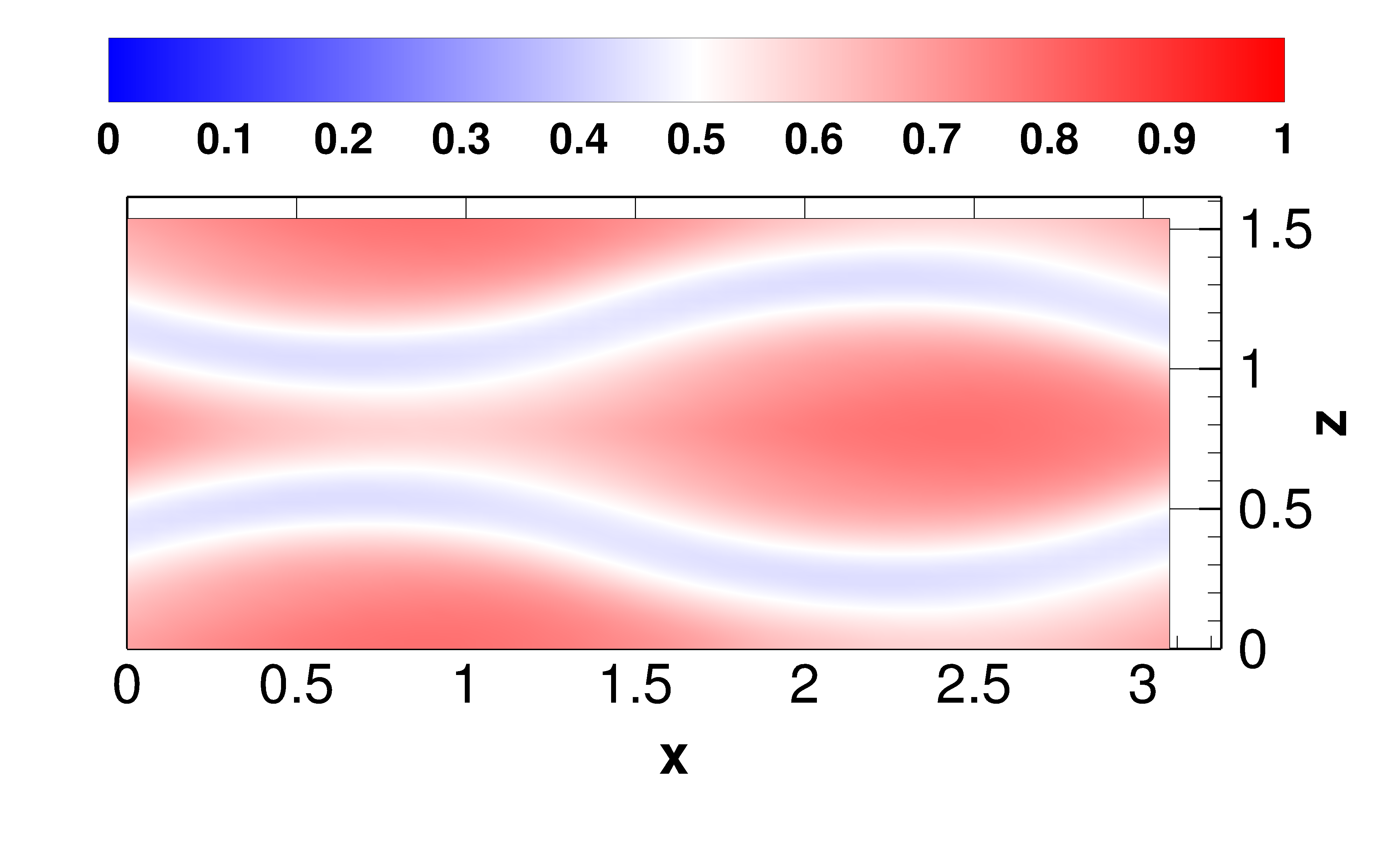} 
  			 \label{fig:LBT666_7}
  			}
  			\caption[Structure of S1 at $\Rey=666$]{(a) Vortical structure of the S1 solution at the saddle node bifurcation at $\Rey=666$ (point \textbf{a} in figure \ref{fig:P4bifdiagram}). Isosurface of $Q=0.05$ is shown with colors indicating distance from wall -- blue near wall, red near center. A color contour plot of streamwise velocity at $x=0.75$ is also shown. (b) Streamwise velocity $u$ at $y^+=20$.}
  			\label{fig:S1SNB}
		\end{center}
		\end{figure}

	Figure \ref{fig:P4bifdiagram} is the bifurcation diagram for the ECS solutions that we have computed in this study, shown in terms of dissipation rate $D$ and streamwise wavespeed $c$ \emph{vs.}~$\Rey$.
	Two solution families are shown. The one of primary interest is shown in purple; it arises in a saddle-node bifurcation at $\Rey=666$ (point \textbf{a} in Figure \ref{fig:P4bifdiagram}).  This solution family satisfies both the $z$-reflection symmetry \eqref{eq:reflect} and the $xz$-shift symmetry \eqref{eq:xzshift} and will be denoted S1. ECS with this set of symmetries have been reported in channel flow by \cite{Wall:2016kw} (their MS-S, MS-A and TW2 families), but in general have received little attention in the literature. The vortex structure of this state at the bifurcation point is illustrated in Figure \ref{fig:S1SNB}a, which shows a surface of vortex strength, $Q=\frac{1}{2}(\|\boldsymbol{\Omega}\|^2 - \|\boldsymbol{\Gamma}\|^2)$ in the bottom half of the channel ($y<0$), where $\boldsymbol{\Omega}$ and $\boldsymbol{\Gamma}$ are the local vorticity and strain rate, respectively and $||\cdot||$ denotes Frobenius norm. This structure consists of two pairs of staggered counterrotating streamwise vortices that are mirror symmetric across $z=L_z/2$. By \eqref{eq:centerplanesym}, a $y-$reflection-symmetric vortex structure arises in the top half of the channel ($y>0$). Figure \ref{fig:LBT666_7} shows the streamwise velocity at $y^+=20$, clearly illustrating the high degree of symmetry of the flow field and the ``sinucose'' nature of the flow pattern -- there is a $z$-reflection-symmetric pair of sinuous low-speed streaks (blue). Observe that the low speed streaks (blue) ``pinch'' together at $x\approx 0.75$. \revision{As mentioned above, \cite{Duguet:2012tw} observe in their study of edge states in boundary layer transition the formation of transient hairpins above pinches. We return to this point below.}
	
\begin{figure}
 		\begin{center}
			\subfigure[]
  			{
 				 \includegraphics[width=0.45\textwidth]{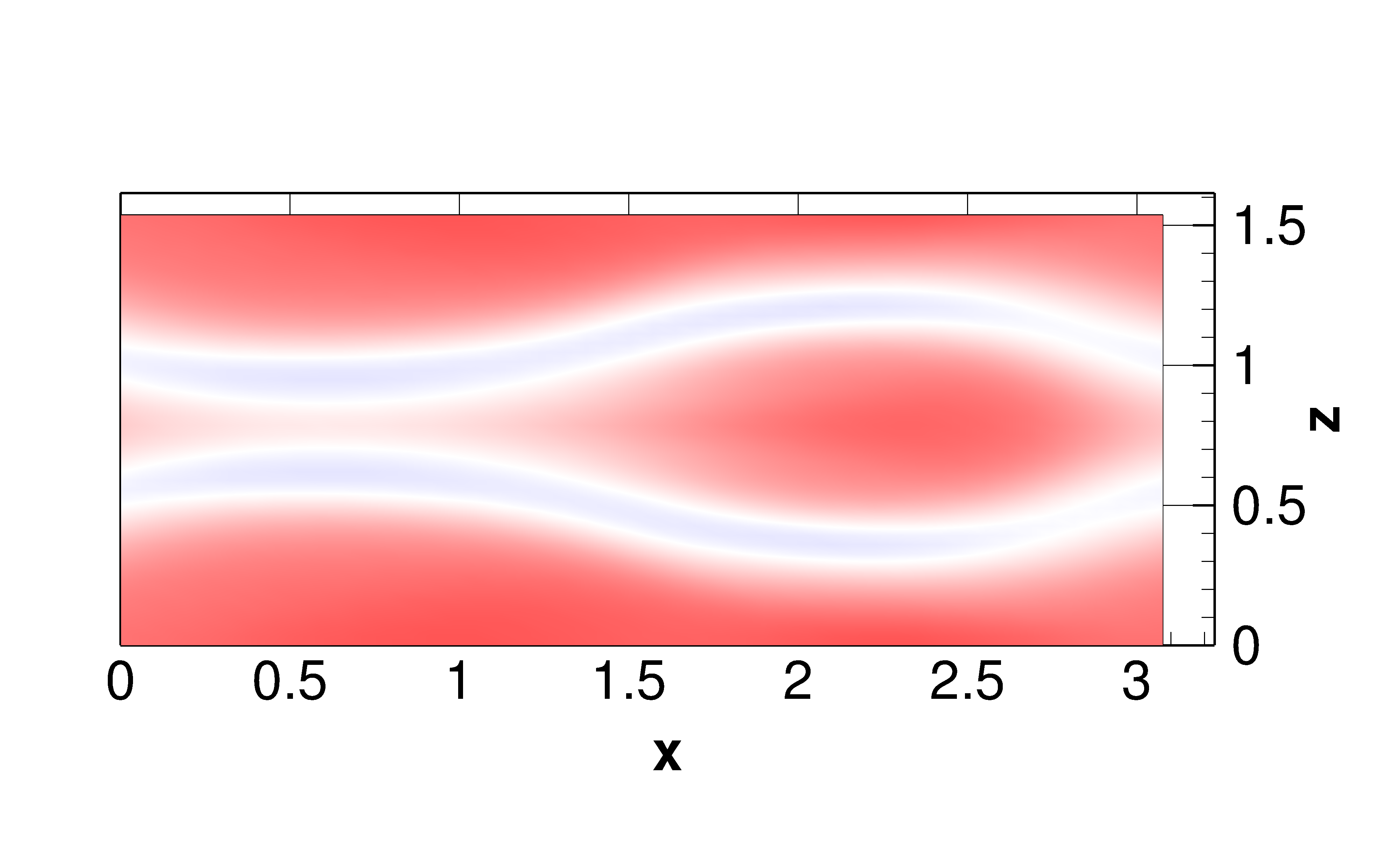} 
  				 \label{fig:UB2_935_2}
 		 	}
 		 	 \hskip 4ex
 			 \subfigure[]
  			{
 			 \includegraphics[width=0.45\textwidth]{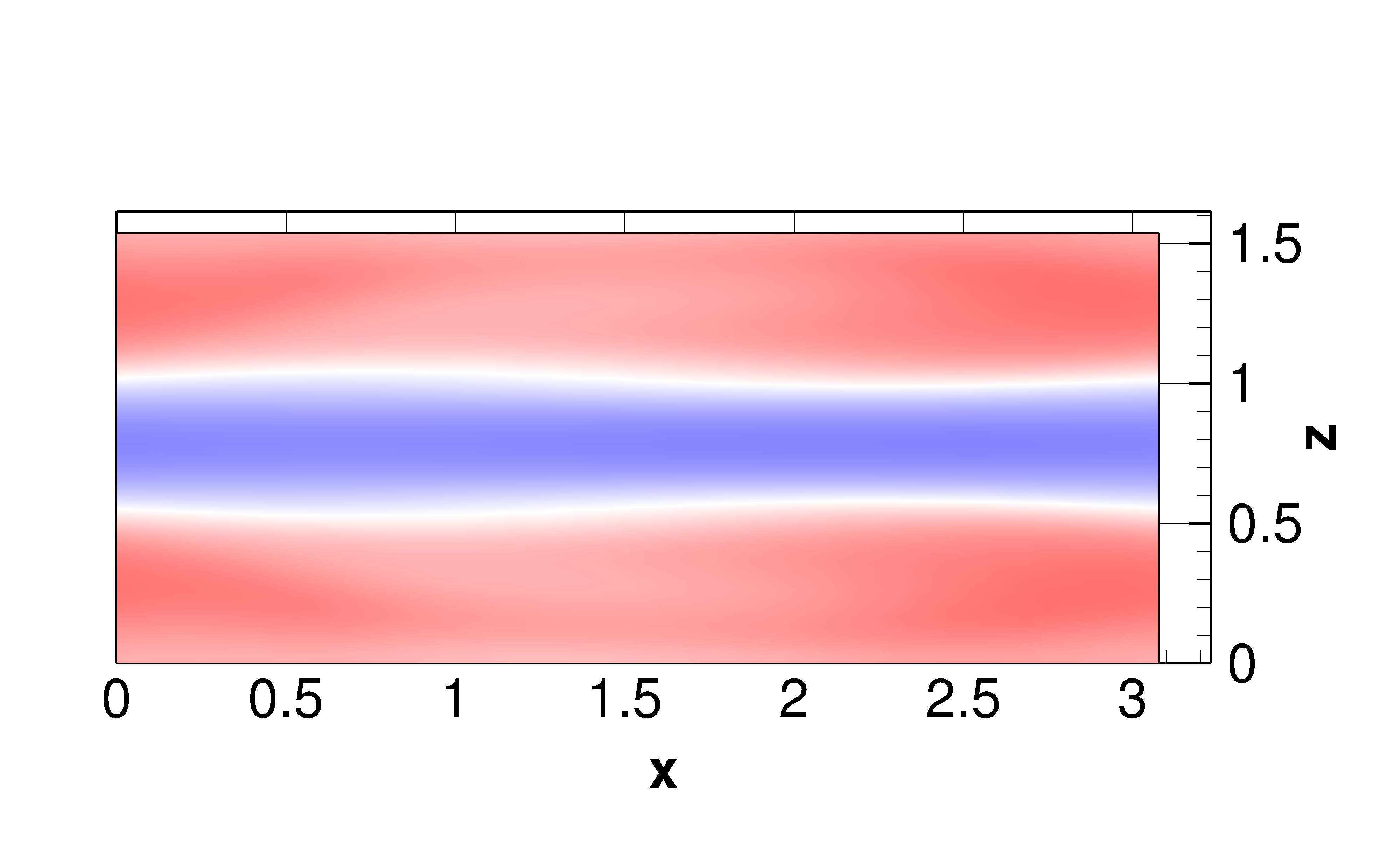} 
  			 \label{fig:UB1_1800_4}
  			}
  			\caption[Streak structure]{Streamwise velocity at $y^+=20$ on P4: (a)  near the pitchfork at $\Rey=935$ (point \textbf{c} in Figure \ref{fig:P4bifdiagram}), (b) at $\Rey=1800$ (point \textbf{d}).}
  			\label{fig:P4streak}
		\end{center}
		\end{figure}

	 On the upper (high dissipation) branch of S1, which we denote S1UB, a pitchfork bifurcation occurs at $\Rey=855$ (point \textbf{b} in Figure \ref{fig:P4bifdiagram}). The bifurcating solutions lack the $xz$-shift symmetry, as shown by the streak structure for a solution on this branch at $\Rey=935$ on Figure \ref{fig:P4streak}a (point \textbf{c} on the bifurcation diagram). This solution family is the P4 family reported by \cite{Park:2014wt}; indeed the way we found S1 was by continuation of P4. As $\Rey$ increases along the P4 solution branch, the two mirror symmetric low-speed streaks merge into one; the structure evolves from sinucose to varicose. Figure \ref{fig:P4streak}b shows this varicose structure at point \textbf{d} on the bifurcation diagram. Upon continuation to higher $\Rey$, P4 turns back on itself several times in saddle-node bifurcations resulting in a rather tangled bifurcation diagram. 

\begin{figure}
 		\begin{center}
			\subfigure[]
  			{
 				 \includegraphics[width=0.45\textwidth]{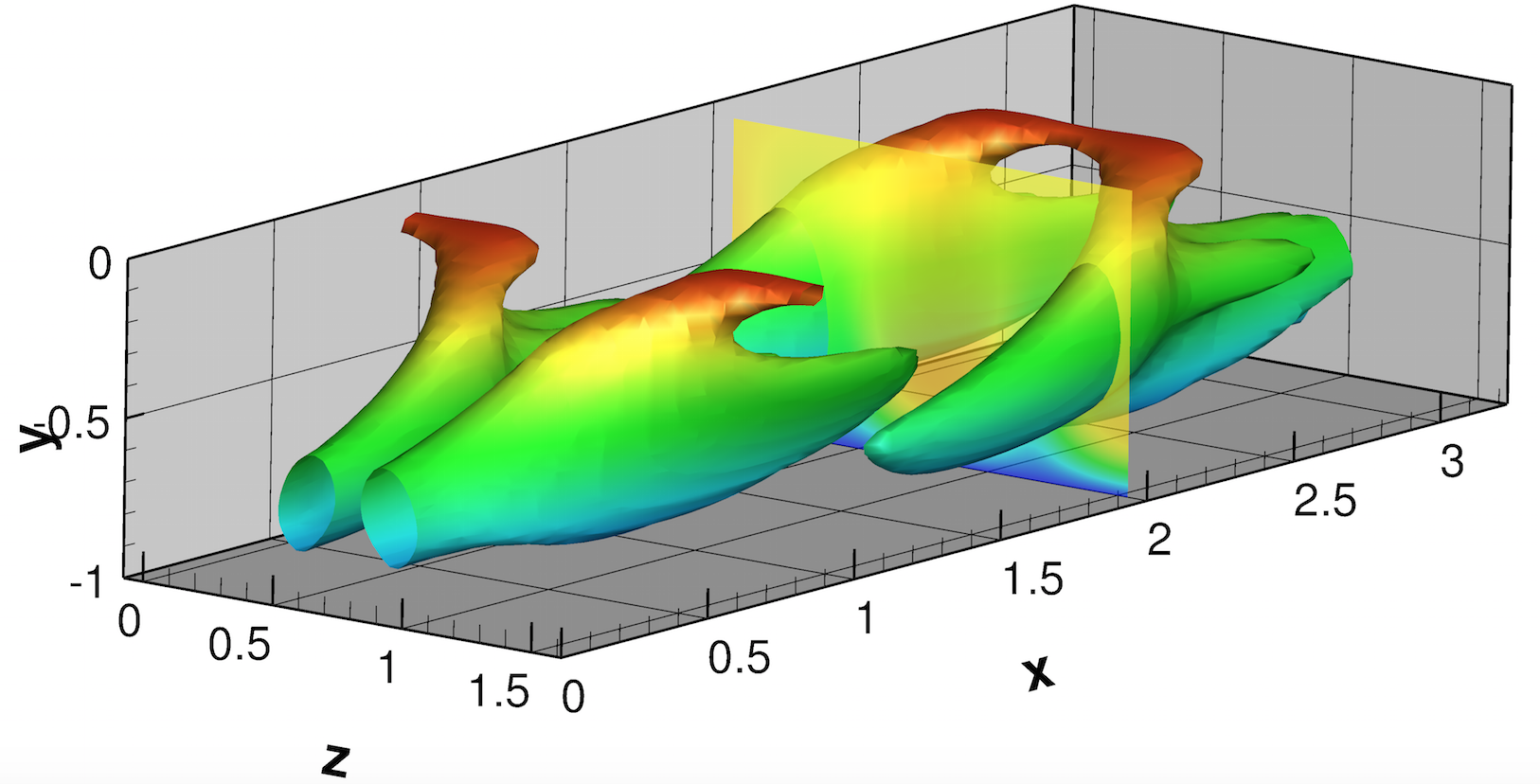} 
  				 \label{fig:UBT_1800_3}
 		 	}
 		 	 \hskip 4ex
 			 \subfigure[]
  			{
 			 \includegraphics[width=0.45\textwidth]{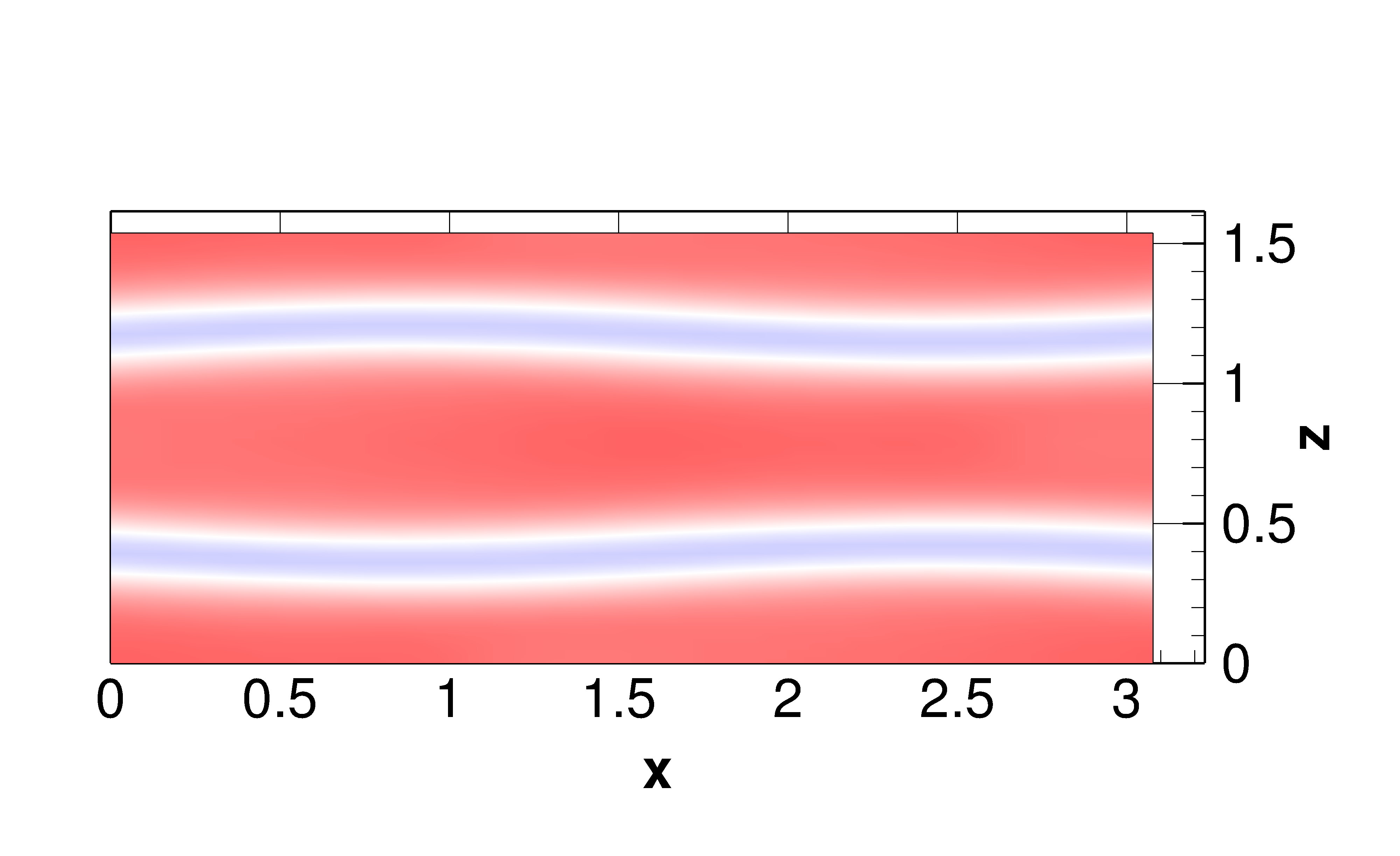} 	
  			 \label{fig:UBT_1800_4}
  			}
  			\subfigure[]
  			{
 			 \includegraphics[width=0.45\textwidth]{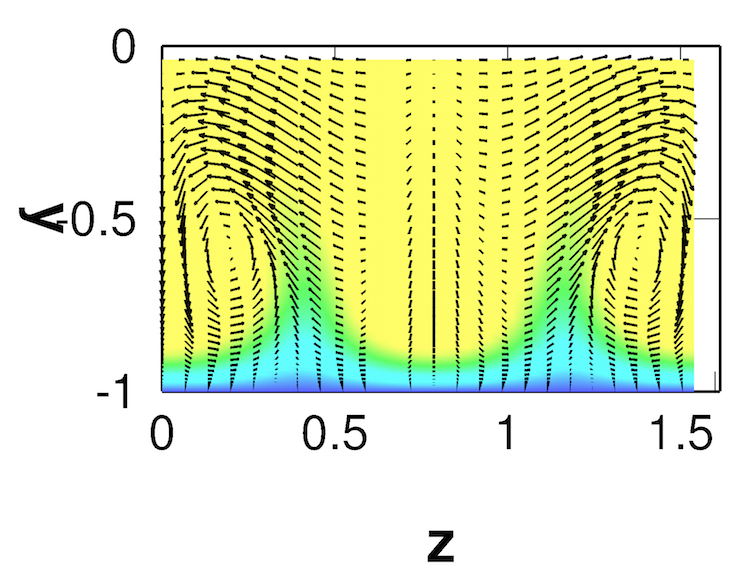} 	
  			 \label{fig:UBTx2}
  			}
  			\caption[Structure of S1UB at $\Rey=1800$]{(a) Structure on S1UB at $\Rey=1800$ (point \textbf{e}). Isosurface of $Q=0.072$ with slice of $u$ at $x=2$ is shown. \revisionb{The maximum value of $Q$ on the $z$ midplane is $Q=0.096$.}(b) Streamwise velocity at $y^+=20$.  (c) Streamwise velocity $u$ at position shown in (a) at $x=2$ with arrows of in-plane velocity vectors.}
  			\label{fig:P4UBT1800}
		\end{center}
		\end{figure}

	Now we return our attention to S1. The upper branch (higher dissipation) solution displays an initial sharp increase in $D$ with distance from the bifurcation point and the structure that emerges on this branch is very interesting and important. Recall the vortical structure of S1 at its inception, as shown in Figure \ref{fig:SNB}. This is a counter-rotating streamwise vortex structure, as expected for near-wall turbulence. By contrast, Figure \ref{fig:UBT_1800_3} shows the vortical structure of S1UB at $\Rey=1800$ (point \textbf{e} on the bifurcation diagram). By this point, the structure has evolved to become a self-sustaining combination of streamwise vortex structure (blue and and green) near the wall and \emph{hairpin\revision{-like}} heads (red) near the centerplane.  It is a microcosm of near-wall turbulence, where streamwise vortices dominate the structure close to the wall, with hairpins becoming dominant further away \citep{WU:2009fn}. \revision{	Here the heads seem to be pulling ahead of the legs a bit, perhaps because of forward motion induced by the presence of a mirror-image hairpin head on the other side of the centerplane $y=0$.} \revision{Consistent with the observations of \cite{Duguet:2012tw}, the hairpins appear above pinches in the streak structure, the most apparent of which is at $x\approx 2.5$. \revisionb{The pinch here is less pronounced than that observed in Figure \ref{fig:LBT666_7}}. By symmetry, there are also pinches at $x\approx 1$ between the low speed streaks in the domain shown and those in the spanwise neighboring domains. Correspondingly, there are hairpin-like structures above these pinches. Half of each of these can be seen in Figure \ref{fig:UBT_1800_3}. Overall, this ECS displays a staggered array of these structures.}  

\begin{figure}
 		\begin{center}
 			 \subfigure[]
  			{
 			 \includegraphics[width=0.6\textwidth]{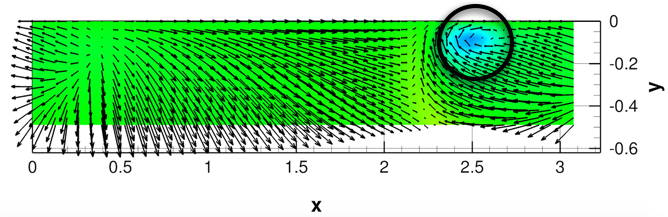} 
  			 \label{fig:ZvortUBT}
  			}
			\subfigure[]
  			{
 				 \includegraphics[width=0.6\textwidth]{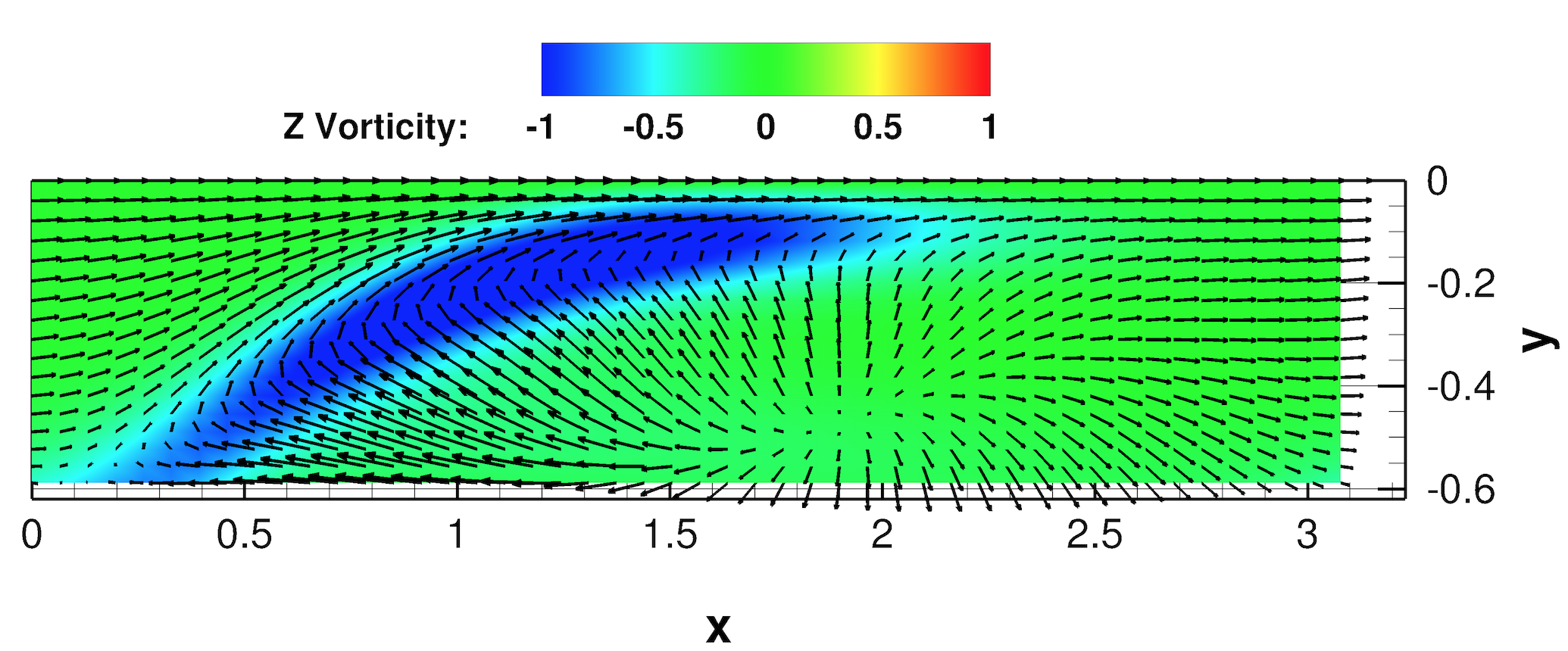} 
  				 \label{fig:ZvortSNB}
 		 	}
  			\caption[Z midplane]{{ Contour of $z$-vorticity on the $z$-midplane with arrows indicating the velocity field in a frame moving at (a) $v_{x} = 0.76$  for S1UB at $\Rey=1800$ and (b) $v_{x} = 0.74$  for S1 at the saddle-node bifurcation point $\Rey=666$. \revisionb{In these plots the region between $y=-0.6$ and $y=0$ (the centerplane) is shown.}}}
  			\label{fig:Zvorticity}
		\end{center}
		\end{figure}

\revision{From Figure \ref{fig:UBT_1800_4}, we can estimate the width of the hairpin-like structure as the spacing between the streaks in the pinch region, which is about $1$ in outer units.  Similarly, from Figure \ref{fig:ZvortUBT} we can estimate the distance of the vortex head from the wall in the same units as $0.9$.  At $\Rey=1800$, the friction Reynolds number $\Rey_\tau\approx 95$. As this quantity gives the scaling between outer and inner units, we see that the vortex structure here has spanwise and wall-normal dimensions of about $95$ and $86$ wall units, respectively. For comparison, \cite{Adrian:2007hy} describes a ``mature'' hairpin vortex packet as having similar size, with both width and height of about 100 wall units.}
	
		

\revision{Further indication of the hairpin-like structure displayed by this ECS is}  illustrated in Figure \ref{fig:ZvortUBT}, which shows the velocity field and $z$-vorticity on the symmetry plane $z=L_z/2$ in a reference frame moving with the transverse vortex core. Toward the upper right of the plot the localized vortical motion corresponding to the hairpin head is clearly seen. It should be emphasized that this motion occurs just below the centerplane of the channel, where the mean shear and thus the mean $z$-vorticity vanish.  This is in sharp contrast to the putative hairpin states described by \cite{Itano:2009ho}; as noted above, those structures were only identified as hairpins based on the shape of the vortex lines, and additionally  the hairpin shape of these lines was observed near the boundaries where there is a strong mean shear.  In contrast, \revisionb{at $\Rey=1800$,} S1UB displays a clear localized region of cross-stream vorticity in a nearly shear-free mean flow that is connected to streamwise vortex motions closer to the wall.  To our knowledge this is the first observation of a persistent hairpin-\revision{like} vortex structure in an exact coherent state. \revisionb{Finally, Fig.~\ref{fig:ZvortSNB} shows the same quantities plotted for the structure at $\Rey=666$, i.~e.~right at the saddle-node bifurcation point. Consistent with the $Q$ structure shown in Fig.~\ref{fig:SNB}, no localization of $z$-vorticity is seen.} 

\begin{figure}
 		\begin{center}
			\subfigure[]
  			{
 				 \includegraphics[width=0.45\textwidth]{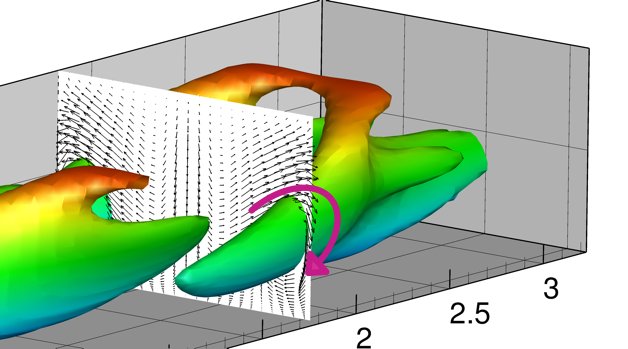} 
  				 \label{fig:Zoom1}
 		 	}
 		 	\hskip 4ex
 			 \subfigure[]
  			{
 			 \includegraphics[width=0.45\textwidth]{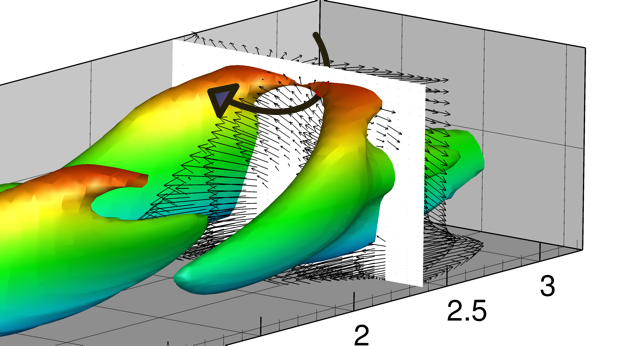} 	
  			 \label{fig:Zoom2}
  			}
  			\subfigure[]
  			{
 			 \includegraphics[width=0.45\textwidth]{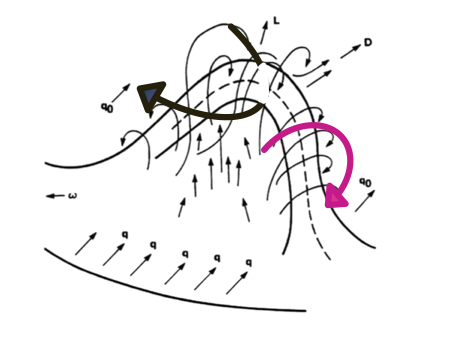} 	
  			 \label{fig:Theo}
  			}
  			\caption[Tying back to Theodorsen]{\revision{(a) In-plane velocity fluctuations on the legs of the vortex structure at $x=1.82$ of S1UB, $\Rey=1800$, (b) Out of plane velocity fluctuations on the head of the vortex structure at $x=2.43$  of S1UB, $\Rey=1800$ and (c) Theodorsen's picture of a hairpin vortex.}}
  			\label{fig:Theodorsen}
		\end{center}
		\end{figure}

\revision{Completing our presentation of the structure of S1UB at $\Rey=1800$, Figures \ref{fig:Zoom1} and \ref{fig:Zoom2} show the sense of rotation of the legs and head of the hairpin-like structure, respectively. Additionally, a strong Q2 ejection event ($u'<0, v'>0$ ) can be seen behind and below the vortex head. For comparison, Figure \ref{fig:Theo} shows again the Theodorsen picture; arrows have been added to all of these figures to indicate that S1UB does indeed display the same qualitative vortex structure as the original Theodorsen hairpin.}

\revision{In addition to the vortex head and leg structure and the Q2 event behind the head just described, \cite{Adrian:2007hy} asserts, based on conditional averaging of the structure associated with Q2 events, that there are two additional ingredients to the signature of a hairpin vortex, a stagnation point separating Q2 and Q4 regions, corresponding to an inclined shear layer behind the hairpin head, and a low-speed streak well behind the vortex. These features are absent from the S1 ECS. Regarding the inclined shear layer and Q2/Q4 stagnation point, this is probably precluded by the reflection symmetry of the channel geometry and the proximity of the vortex heads to the centerplane. Regarding the trailing low speed streak, this is precluded by the symmetries \eqref{eq:reflect} and \eqref{eq:xzshift}, which yield a staggered arrangement of hairpin-like vortices.} \revisionb{It should also be emphasized that the ECS structures described here are not conditionally-averaged -- it is not obvious that instantaneous hairpin-like structures in flow will display all the features of a conditionally-averaged one. Nevertheless, it would be interesting to learn whether ECS exist that have aligned rather than staggered hairpin-like structures --these are likely to bear more similarity to the hairpins described by Adrian than is found here.}

\revision{Nevertheless, as noted above, the scale of the S1 ECS vortex structure is similar to that reported by \cite{Adrian:2007hy}. Furthermore, it is consistent with observations of hairpins in the transition context as exemplified by the observations of \cite{Duguet:2012tw}, where hairpins arise above the pinch region between two low speed streaks. Perhaps the S1 ECS is more related and relevant to hairpins observed in transition than to those observed in fully turbulent boundary layers.}

		\begin{figure}
 		\begin{center}
			\subfigure[]
  			{
 				 \includegraphics[width=0.45\textwidth]{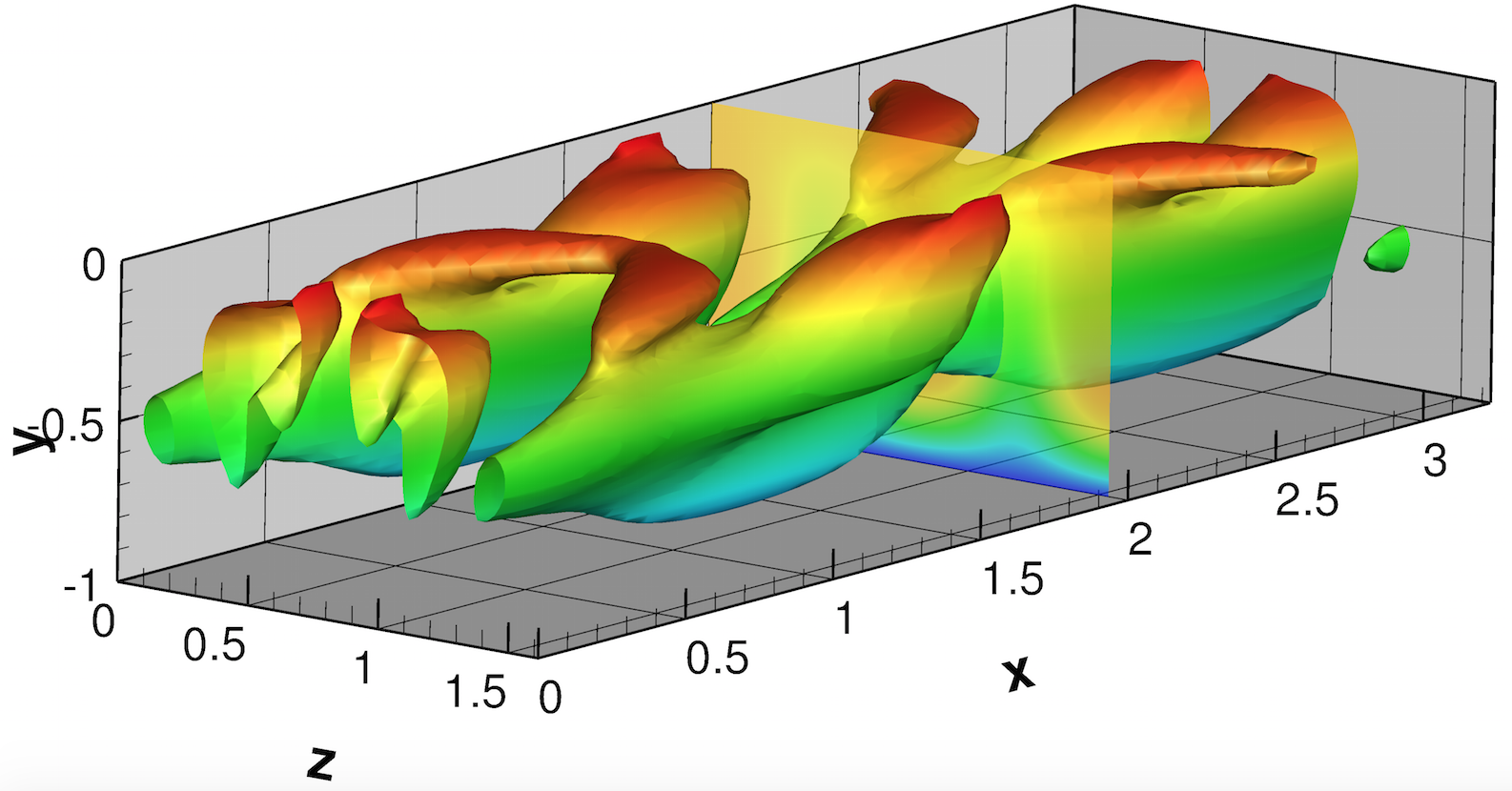}
  				 \label{fig:LBT_3000_5}
 		 	}
 		 	 \hskip 4ex
 			 \subfigure[]
  			{
 			 \includegraphics[width=0.45\textwidth]{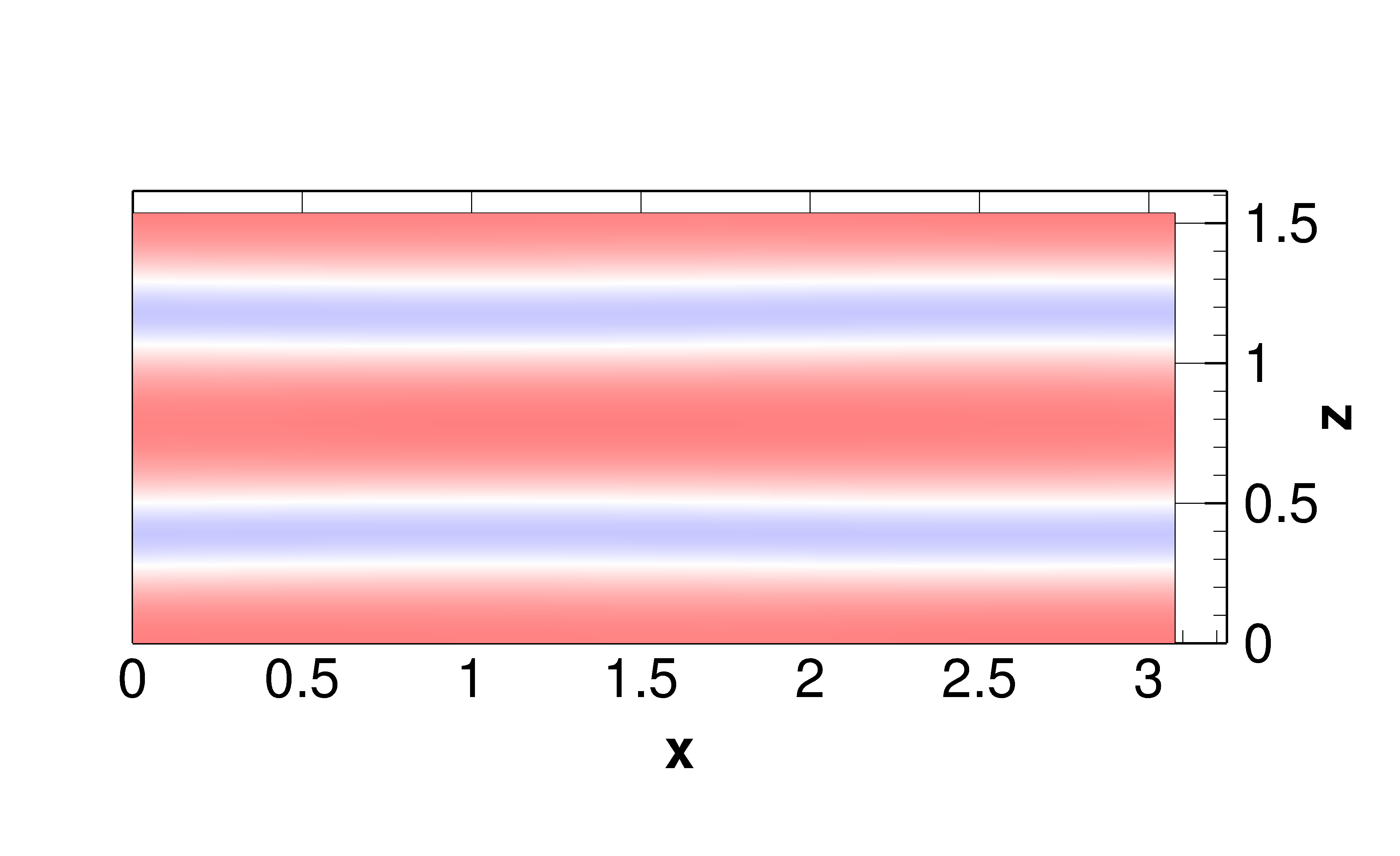}
  			 \label{fig:LBT_3000_6}
  			}
  			\caption[Structure of LBT at $\Rey=3000$]{(a) Vortical structure of S1LB at $\Rey=3000$ (point \textbf{i}). Isosurface of $Q=0.014$ is shown. (b) Streamwise velocity $v_{x}$ at $y^{+}=20$.}
  			\label{fig:P4LBT3000}
		\end{center}
		\end{figure}
		
		\begin{figure}
 		\centering
 		\includegraphics[width=0.7\textwidth]{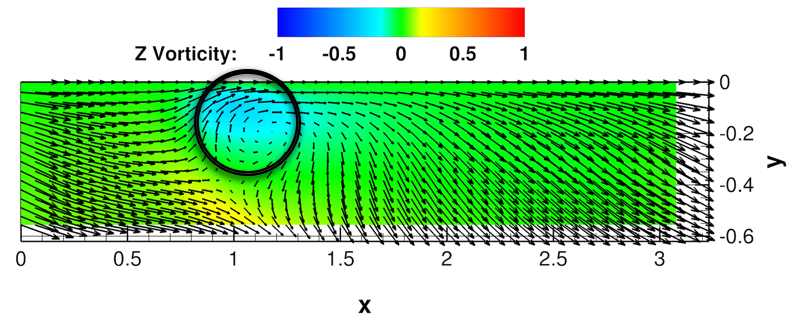}
  		\caption[Z midplane]{Contour of $z$-vorticity on the $z$-midplane with arrows indicating the velocity field in a frame moving at $v_{x} = 0.775$ for S1LB at $\Rey=3000$. \revisionb{In this plot the region between $y=-0.6$ and $y=0$ (the centerplane) is shown.}}
  		\label{fig:ZvortLBT}
		\end{figure}
		
		\begin{figure}
		\begin{center}
			\subfigure[]
  			{
 				 \includegraphics[width=0.44\textwidth]{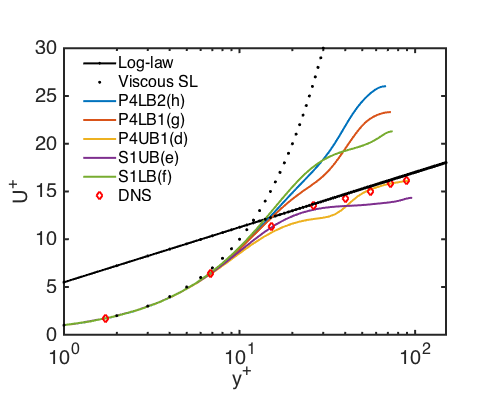}
  				 \label{fig:umean}
 		 	}
 		 	 \hskip 0.5ex
 			 \subfigure[]
  			{
 			 \includegraphics[width=0.44\textwidth]{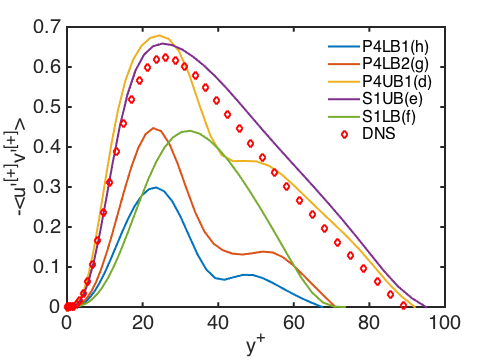}
  			 \label{fig:Rssplus}
  			}
  			\subfigure[]
  			{
 				 \includegraphics[width=0.44\textwidth]{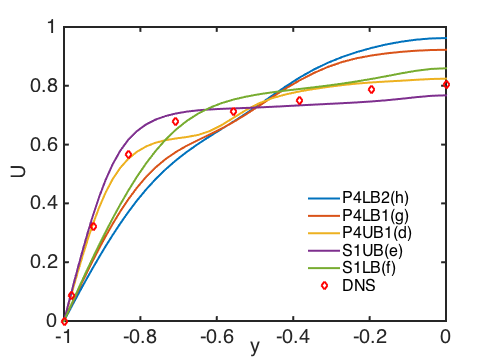}
  				 \label{fig:UmeanOuter}
 		 	}
 		 	 \hskip 0.5ex
 			 \subfigure[]
  			{
 			 \includegraphics[width=0.44\textwidth]{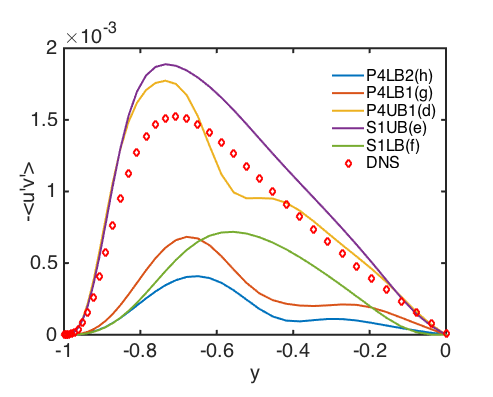}
  			 \label{fig:rssmean}
  			}
  			\caption[Profiles]{Mean profiles for ECS and turbulent flow (DNS) at $\Rey=1800$: (a) Mean velocity in inner units, (b) Reynolds shear stress in inner units, (c) Mean velocity in outer units and (d) Reynolds shear stress in outer units. For the ECS, the parenthesized letter in the legend indicates the corresponding point on Figure \ref{fig:P4bifdiagram}.}
  			\label{fig:profiles}
		\end{center}
		\end{figure}

\revision{In prior observations, hairpin vortices have been observed as transient, spatiotemporally evolving structures. In contrast, the hairpin-like ECS structure observed in Figure \ref{fig:P4UBT1800} is, in a traveling reference frame, an equilibrium structure.  We believe that the reason it was possible to find such a structure is the channel geometry and the corresponding centerplane symmetry; this keeps the hairpin from continually drifting away from the wall as occurs in a boundary layer -- the channel geometry has allowed us to ``trap'' the hairpin structure, confining it to a finite distance from the wall and allowing it to propagate as a traveling wave. An interesting topic of future work would be to seek ECS with streamwise aligned, rather than staggered, hairpin-like vortices.}
 
The above discussion has focused on the evolution of S1 from the bifurcation point in the direction of increasing dissipation. \revisionb{For completeness, we show in Figures \ref{fig:P4LBT3000} and \ref{fig:ZvortLBT} the structure that evolves along the lower branch solution S1LB.
Characteristic of lower branch solutions, the overall vortex and streak structure becomes much weaker with distance from the saddle-node bifurcation point, and while a $z$-oriented vortex structure also emerges on this branch, it is very weak, and not well-defined until higher $\Rey$ than on the upper branch. This is why we have illustrated the structure at $\Rey=3000$ rather than the value of $1800$ that we used for the upper branch. Observe that that the $Q$ value chosen for Fig.~\ref{fig:LBT_3000_5} is 0.014, much smaller than the value 0.072 used to visualize the upper branch (Figure \ref{fig:UBT_1800_3}) and the waviness of the streaks shown in Fig.~\ref{fig:LBT_3000_6} is very weak -- they are very nearly streamwise invariant.}


An important issue is the relationship between ECS and turbulent dynamics.  Figure \ref{fig:profiles} shows profiles of area-averaged mean velocity and Reynolds shear stress at $\Rey=1800$ for the states indicated with red dots on Figure \ref{fig:P4bifdiagram}; these include solutions from both the P4 and S1 families.  Also shown are profiles from turbulent trajectories obtained via direct numerical simulation (DNS) for the same domain and Reynolds number. Both S1UB and P4UB1 have mean velocity profiles that are fairly close to that of the turbulence, though the latter has inflectional kinks that are absent in S1UB.  In inner units, the Reynolds shear stress for S1UB is in quantitative agreement with the turbulent result near the wall and yields very nearly the same peak position $y^+\approx 25$. For $y^+\gtrsim 20$, S1UB has slightly higher Reynolds shear stress than the turbulence. In outer units, the quantitative resemblance is not as strong, but is still reasonably good. In any case, S1UB more closely resembles the turbulence than do any of the other ECS reported here.

\begin{figure}
 		 \begin{center}
 			 \subfigure[]
  			{
 				 \includegraphics[width=0.45\textwidth]{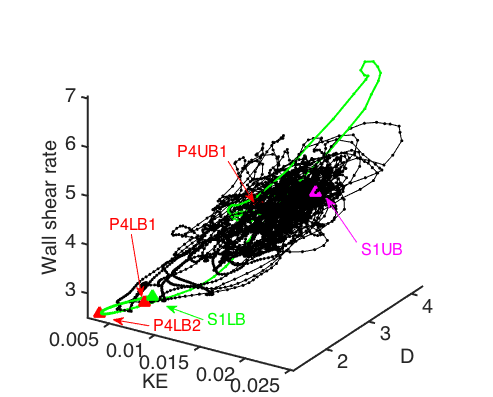}
 				 \label{fig:ss18_1}
  			}
  			 \hskip 0.5ex
  			 \subfigure[]
 		    {
  				\includegraphics[width=0.45\textwidth]{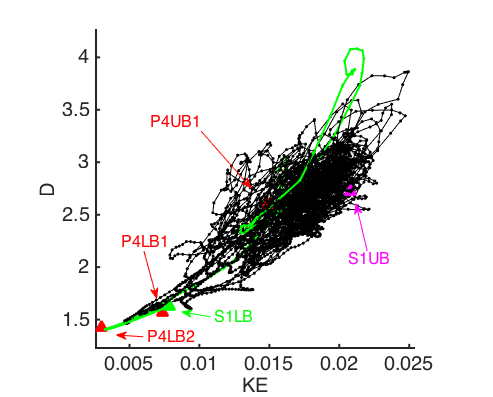}
 				\label{fig:ss18_2}
  			}
  			\subfigure[]
  			{
 			 \includegraphics[width=0.4\textwidth]{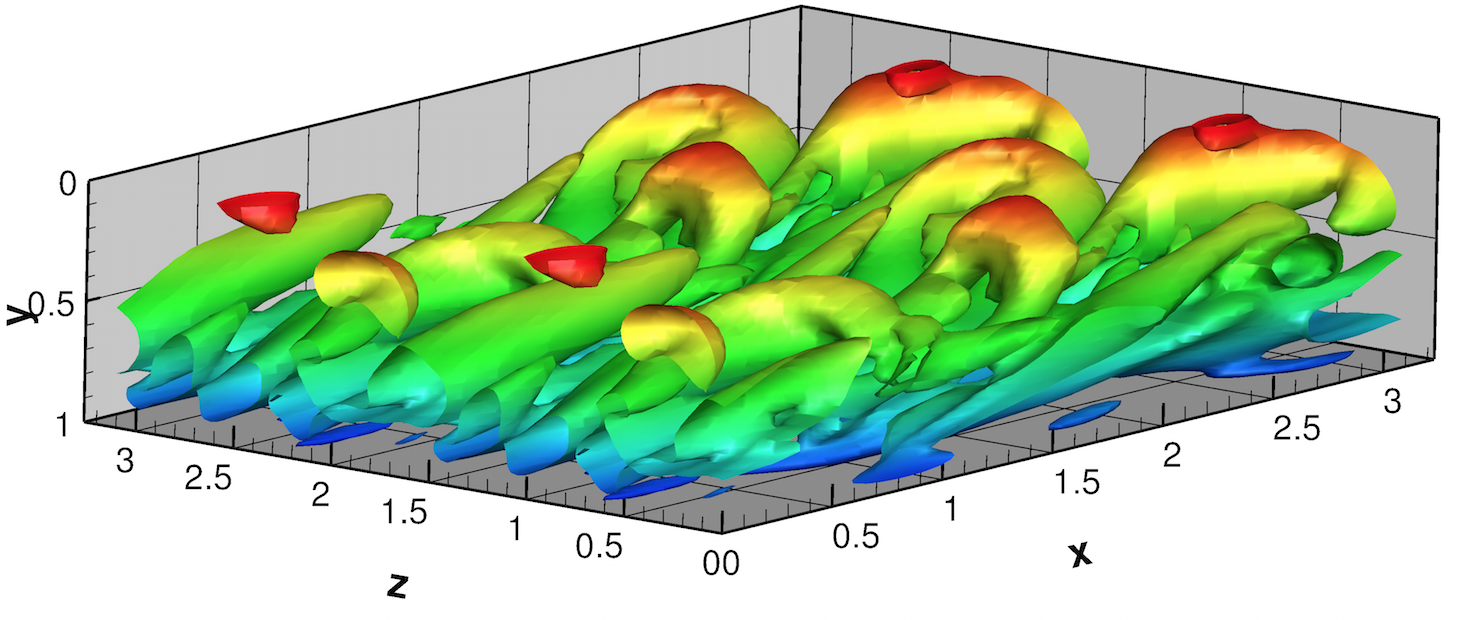}
  			 \label{fig:Hairpin_MFUp072}
  			}
		\caption[State space projection of turbulent dynamics at $\Rey  =  1800$]{(a) State space visualization of DNS trajectories at $\Rey=1800$, projected onto three dimensions: disturbance kinetic energy(KE), energy dissipation rate(D) and wall shear rate. The labelled symbols are computed ECS and the black lines are the DNS trajectories at $\Rey=1800$. (b) 2D projection of (a). Also shown is a trajectory starting from S1LB (green). 
		\revision{(c) Isosurface of $Q=0.072$ at an instant during DNS at $\Rey=1800$. Shown is the top half of the domain, rotated to be consistent with other figures. For clarity, two periods in $z$ are shown.}
		}
  		\label{fig:StateSpace}
		\end{center}
		\end{figure}

Another illustration of the relationship between ECS and turbulence is given in Figure \ref{fig:StateSpace}, which shows projections of ECS and trajectories onto selected variables:  area-averaged wall shear rate, dissipation rate $D$ and disturbance kinetic energy $KE=\frac{1}{2}||\boldsymbol{u}-\boldsymbol{u}_{\textrm{laminar}}||^{2}$. In this projection, all of the ECS plotted appear to be embedded in the same region of state space as the turbulent attractor. Also shown in green in Figure \ref{fig:StateSpace} is a trajectory starting from S1LB plus a random perturbation. It starts to approach P4LB2 before bursting towards the turbulent core. This observation points at a possible heteroclinic connection between the S1LB and P4LB2 and at a role that hairpins may play in the transition to turbulence. 

\revision{Finally, while hairpin structures are not prominent in DNS under the conditions considered here, they do arise occasionally, as seen in the simulation snapshot shown in Figure \ref{fig:Hairpin_MFUp072}.}

%% file: Conclusions.tex
Prior studies have created a catalogue of nonturbulent recurrent solutions to the Navier--Stokes equations  -- exact coherent states -- that capture key features of near-wall turbulence, including  streamwise vortices \citep{Waleffe:2001wu}, bursts \citep{Toh:2003df} and localized spots \citep{Brand:2014ww} or puffs \citep{Chantry:2014ex}. 
The present work adds an important entry to this catalogue. \revision{Hairpin vortices have long been observed within turbulent flows in various contexts, from boundary layer transition to fully-developed flows. }  Both the upper and lower branches of the S1 solution family reported here for the minimal channel flow geometry display a clear hairpin\revision{-like} structure, with a spanwise-oriented head connected to quasistreamwise-oriented legs. 
Additionally, in the Reynolds number regime studied, the upper branch state has mean velocity and Reynolds shear-stress profiles that are quantitatively similar to the turbulent profiles in a channel of the same (minimal) dimensions. 

The present results are of course obtained for a system that is highly constrained by symmetry and periodicity conditions. Nevertheless, the observations described here further strengthen the growing body of work connecting turbulent dynamics to simpler underlying state-space structures. By understanding the latter, we are increasingly gaining insight into the former.